\documentstyle[sprocl]{article}

\newcommand{\beq}{\begin{equation}}   
\newcommand{\eeq}{\end{equation}}

\begin{document}

\title{SUPERSYMMETRIC DOMAIN WALLS 
(NEW  EXACT RESULTS IN SUPERSYMMETRIC GLUODYNAMICS)}

\author{M.  SHIFMAN}

\address{Theoretical Physics Institute, Univ. of Minnesota,
Minneapolis, MN 55455, USA}

\maketitle\abstracts{Supersymmetric domain walls are discussed.
The main emphasis is made on saturated walls in strongly coupled gauge 
theories. The central charge of the ${\cal N} =1$ superalgebra appears
in supersymmetric gluodynamics
as a quantum anomaly. I also consider some rather unusual general 
features of supersymmetric domain walls.}

The subject of my talk today lies somewhat outside the 
mainstream research in supersymmetry (SUSY). On  one hand, it 
has little to do with the supersymmetric phenomenology, and, on the 
other hand,  it  does not  blend with the current ``high theory" -- 
$D$-branes -- although in a couple of points  there is a certain overlap. 
Nevertheless,  I think that the supersymmetric domain wall is a 
beautiful object  that deserves thorough investigation. As we will see, 
this is a   novel theoretical construction which admits exact results
in strongly coupled theories, such as supersymmetric gluodynamics,
 the theory of gluons and gluinos,  the closest 
relative of QCD.  It is assumed that color confinement takes place in 
the strong coupling regime,  much in the same way as in QCD. Usually, 
it is believed that under the circumstances no interesting dynamical 
quantity can be  calculated exactly. I will explain that the domain 
wall tension is {\em exactly} calculable. Although my prime concern is 
SUSY
gluodynamics, some of the results I am going to report will be of a
more general  nature. They refer to any SUSY theory with multiple 
domain walls.

When G. Dvali and I started working on the supersymmetric domain 
walls at the  end of  1996  \cite{DS1,DS2}, I could hardly 
anticipate  that the project would grow and branch to the extent it 
actually did. Quite a few   papers were published \cite{DS1} -- 
\cite{KKS}, and more are in progress.  It goes without saying that it is 
impossible to  present a full account of all the results obtained. I will 
limit myself to main findings; they are summarized below:

$\bullet$ It turns out that in ${\cal N}=1$ theories, SUSY algebras 
may have central extensions. This assertion by itself is  heresy.  If 
you open any text book on supersymmetry you will read that two 
theorems forbid central extensions  in ${\cal N}=1$ superalgebras. 
I will explain how these theorems are circumvented in those cases 
in which domain walls are formed. 

$\bullet$ A zoo of various domain walls is typical of SUSY theories
possessing a discrete set of vacua. I keep in mind that
supersymmetry guarantees the vanishing of the vacuum energy density, 
so multiple degenerate (and physically inequivalent) vacua are quite 
typical.

$\bullet$ In certain domain walls, 1/2 of the supersymmetry is 
spontaneously broken and 1/2 preserved. These are the so called 
BPS walls (more exactly, they should be called BPS-saturated).
The exact results mentioned at the beginning refer to the BPS 
walls. For such walls the energy density per unit area (the tension)
is exactly calculable: it is expressed as the difference of the values of 
the superpotential
far to the left and far to the right from the wall.

$\bullet$ In supersymmetric gluodynamics the role of the
superpotential  is played by the gluino condensate.
This is not seen at the classical level. The central charge in SUSY 
gluodynamics vanishes at the classical level; it emerges
at the quantum level as a quantum anomaly.
The wall tension $\varepsilon$ is
\begin{equation}
\varepsilon = \frac{N_c}{8\pi^2}|\Delta (\mbox{Tr}\, 
\lambda\lambda)|\, ,
\label{wtsg}
\end{equation}
where $N_c$ is the number of colors and $\lambda$ is the gluino 
field. The internal structure of the domain wall may be incredibly
complicated. Still, its tension is given 
by this very simple expression. I should add that the gluino 
condensate, in turn, was exactly calculated
in 1988 \cite{SV1}. 
The fact that the gluino condensate is akin to the superpotential 
was known previously  in a different context \cite{Nati1}. 
The correspondence gets an independent confirmation with the 
calculation of the central charge in SUSY gluodynamics. The 
central charge turns out 
to be proportional to $\lambda\lambda$.

$\bullet$ For certain domain walls there exists an exact degeneracy:
say,
\beq
\varepsilon_{13} = \varepsilon_{12} + \varepsilon_{23}
\label{degen}
\eeq
where $\varepsilon_{ij}$ is the tension of the wall interpolating 
between the  vacuum $i$ and  $j$, respectively. In other words,
if one puts the two domain walls indicated on the right-hand side at an
arbitrary distance from each other, they absolutely do not interact,  
either classically or quantum mechanically, there is neither 
repulsion nor attraction.  I believe that previously 
such domain walls were not known (although in some exactly 
solvable  two-dimensional models a similar property was known for 
kinks). 

After this brief ``executive summary" I will move on to a more 
systematic 
discussion of the 
issues. First of all, how  can one understand the existence of the 
central extensions  of  ${\cal N} = 1$ superalgebras?
A  heuristic explanation was suggested by Witten 
\cite{EW}. Assume  a four-dimensional theory admits domain walls 
(or strings). Compactify the theory in two directions parallel to the 
wall on a very large two-torus of area $A$. The theory then reduces 
to an effective two-dimensional theory where the central extension 
obviously exists \cite{OW}; the central charge for the states 
interpolating between distinct vacua at spatial infinities is
the superpotential difference between the two vacua.  
On the other hand, in the limit $A\to\infty$ we return back to the 
original four-dimensional theory.

Thus, there is 
no central charge for particles in four-dimensional 
${\cal N } =1$ theories, but either strings or domain walls can be 
BPS-saturated (i.e. have their mass equal to the central charge). 

What allows one to circumvent the classic analysis of Haag {\em et 
al.} \cite{HLS} asserting the impossibility of the central extension in 
${\cal N }= 1$ theories? Implicit in this analysis is the assumption
that the full four-dimensional Lorentz invariance is unbroken.
Needless to say that in the sector with the given domain wall
the Lorentz symmetry is spontaneously broken. The bosonic zero 
mode on the wall is the manifestation of this breaking.
It is only the three-dimensional Lorentz invariance that survives
(one time plus two spatial directions parallel to the wall).

The actual calculation of the central charge is quite trivial in the 
generalized Wess-Zumino models where it appears at the tree level
\cite{DS1,CS}. One calculates $\{ Q_\alpha Q_\beta\}$ using
the canonic commutation relations. Terms that are spatial integrals of 
total derivatives, that are usually dropped, must be kept, since
the central charge emerges as a spatial integral of a total derivative.
This is the only relatively subtle point. The result is  not worth 
presenting here since shortly I will give a more  general formula 
valid in gauge theories with arbitrary superpotentials.

The basis of any gauge theory is  supersymmetric gluodynamics, and its 
Lagrangian  is 
\beq
{\cal L} = \left\{\frac{1}{4 g_0^2}{\rm Tr}\;\int d^2\theta\;\; W^2\; + 
\mbox{H.c.}\right\} = \frac{1}{g_0^2}\left\{ - {\frac{1}{4 
}G_{\mu\nu}^aG_{\mu\nu}^a
+i\bar\lambda^a_{\dot\alpha}{\cal 
D}^{\dot\alpha\alpha}\lambda^a_\alpha}\right\} \, .
\label{SGD}
\eeq
Note that in my normalization the 
coupling constant is an overall factor in front of Tr$W^2$.
For simplicity I will limit myself to SU($N$) gauge groups, although 
the analysis can be readily made general.

Superficially, this theory is very similar to QCD -- the only distinction 
is the representation of the fermion fields. In QCD the fermion fields
belong to the fundamental representation, while in 
supersymmetric gluodynamics they are the Weyl fields in the adjoint 
representation. The theory is strongly coupled. It is believed that
the theory confines color more or less in the same way as  QCD.
I hasten to add that, perhaps, this is not quite true, as will be 
explained shortly.

Supersymmetric gluodynamics
has multiple discrete vacua. The Lagrangian (\ref{SGD})
is invariant under U(1) which is broken by the anomaly down to 
$Z_{2N}$. The expectation value
$\langle \lambda^2\rangle$ is the 
order parameter; the gluino condensation implies that
the discrete chiral symmetry $Z_{2N}$
is spontaneously broken down to $Z_2$,
\beq
\langle\lambda^2\rangle = \Lambda^3\exp \left( 2\pi i k / N
\right)\, ,\,\,\,  k = 0, 1, 2, ...., N-1\, .
\label{gc}
\eeq
 The gluino condensation was 
conjectured by Witten in the early 1980's \cite{WI}.
The value of the gluino condensate was exactly calculated \cite{SV1}
by adding matter in an appropriate representation, Higgsing the 
theory -- to enforce  the weak coupling regime --
and then returning back to SUSY gluodynamics by exploiting 
holomorphy. 

The number of vacua labeled by the gluino condensate (\ref{gc}) is
$N$. This is precisely the value of Witten's index for the
SU($N$) SUSY gluodynamics. Thus, the chirally asymmetric vacua 
saturate Witten's index. Recently it was argued \cite{KS}
that the theory admits also a chirally symmetric vacuum 
($\langle\lambda^2\rangle=0$) in which no mass gap develops. In 
this vacuum, the theory is presumably in the conformal regime; the
energy density becomes positive in finite volume, so that the 
chirally symmetric vacuum does not contribute to Witten's index. It 
plays a role, however, 
in the domain wall dynamics.

Thus, we have multiple vacua which are degenerate. Supersymmetry 
guarantees that the vacuum energy density vanishes in each of these 
states. If so, there must exist field configurations interpolating 
between  distinct vacua at infinities. By analogy with solid state 
physics,
one calls them  the domain walls. 
Assume the domain wall
lies in the $xy$ plane. At $z=-\infty$ the order parameter
takes some value, say, $\lambda^2 =-\Lambda^3$.
The value of $\lambda^2$ rapidly changes in the transition domain 
near $z=0$ (the width of the transition domain is of order 
$\Lambda^{-1}$), then $\lambda^2$ approaches $+\Lambda^3$
 at $z=+\infty$. Inside the transition domain 
the fields have to restructure themselves -- there are gradient terms, 
and potential terms -- so that the vacuum energy density
is distorted and is of order $\Lambda^4$. The total  
 wall mass is proportional to $A\Lambda^3$, where $A$
is the wall area, $A\to\infty$.  The wall tension $\varepsilon$ (i.e. 
the energy per unit area) is finite. Since the theory is strongly 
coupled we have no idea how to describe the internal structure of 
the   wall. Yet, the tension is calculable.
If you wish, the wall is a dynamic realization of 2-branes that are so 
popular now.
The $D$-branes are abstract creatures, however, while the walls I am
speaking about are made from ``flesh and blood". 

In order to explain how one can exactly calculate $\varepsilon$,
I must  take a step back and discuss the superalgebra.
The standard ${\cal N} =1 $ superalgebra includes two basic 
anticommutation relations,
\beq
\{Q_\alpha \bar{Q}_{\dot\beta} \} = 2 P_{\alpha\dot\beta }\,  \,\,\, 
\mbox{and} \,\,\,  \{
\bar{Q}_{\dot\alpha} \bar{Q}_{\dot\beta} \} = 0\, .
\label{dop}
\eeq
The vanishing of the anticommutator $\{\bar{Q}
\bar{Q}\} $ 
is a  consequence of the Lorentz symmetry and the Coleman-Mandula 
theorem \cite{CM}. At the classical level Eq. (\ref{dop}) {\em is valid} 
in 
SUSY gluodynamics, as can be seen from  a straightforward 
calculation 
of the canonic commutators. However, at the loop level
a quantum anomaly modifies this relation. In a generic SUSY gauge 
theory with a superpotential ${\cal W}$ and a set of the matter fields 
in the
representations $R_i$ of the gauge group,
one has \cite{DS2,KSS,CS}
\begin{eqnarray}
&&\{\bar{Q_{\dot\alpha}}\bar{Q_{\dot\beta}}\} = \frac{4}{3} 
(\vec{\sigma})_{\dot\alpha\dot\beta} \, \int d^3x  
\vec{\nabla}
\left\{
\left[ 3{\cal W } - \sum_i Q_i \frac{\partial{\cal W }}{\partial Q_i}   
\right]-
\right.\nonumber \\
&&\left.\qquad\quad\qquad
\left[ \frac{3T(G)- \sum_i T(R_i)}{16\pi^2}{\rm Tr}\, W^2 + 
\frac{1}{8}\sum_i\gamma_i 
\bar{D}^2
(\bar{Q}_i^+ e^{V} Q_i) \right]
\right\}_{\theta =0}
\, .
\label{mlpce}
\end{eqnarray}
Here $(\vec{\sigma})_{\dot\alpha\dot\beta}$
is a triplet of matrices transforming the vectorial index of
the  $\{0 , 1\}$
representation of the Lorentz group into the spinorial indices,
$
(\vec{\sigma})_{\dot\alpha\dot\beta} = \{-\tau_3, i , \tau_1  
\}_{\dot\alpha\dot\beta}
$. Moreover, 
$T(R)$ is (one half) of the Dynkin index for representation $R$,
$T(G)$ is (one half) of the Dynkin index in the adjoint  representation
($T(G) = N$ for SU($N$)), $\gamma_i$ are the anomalous dimensions 
of the
matter 
fields $Q_i$. 

The central extension in the right-hand side of Eq. (\ref{mlpce})
is proportional to the spatial integral of a total derivative.
Thus, it vanishes for all localized field configurations
(particles!), in full accord with the classic theorems.
It does not vanish, however, for  wall-like configurations. 
In supersymmetric gluodynamics the central charge is proportional
to the difference in the expectation values of the gluino condensate
in the vacua between which the given wall interpolates
(see Eq. (\ref{wtsg})). When matter fields are added,
the superpotential also enters the game,
in a well-defined manner. If one drops total superderivatives
in Eq. (\ref{mlpce})
(they have no impact on the central charges)
the right-hand side can be rewritten as
\beq
4 (\vec{\sigma})_{\dot\alpha\dot\beta} \int 
d^3x  
\vec{\nabla}
\left\{
{\cal W } - \frac{T(G)- \sum_{i} T(R_i)}{16\pi^2}{\rm Tr}W^2   
 \right\}\, . 
\label{mlpcedop}
\eeq
One can dub the combination in the braces
a {\em generalized superpotential}. That's what replaces superpotential 
when the anomaly is switched on. 

Do we deal  with a new anomaly?
Taking into account the geometric nature of the supercharges, the
occurrence of a new anomaly seems highly unlikely. And, indeed, one 
can show \cite{CS} that the right-hand side of Eq. (\ref{mlpce})
is in one-to-one correspondence with well-known anomaly 
\cite{anom1} in the superdivergence of the supercurrent 
\cite{anom2}, $\bar{D}^{\dot\alpha} J_{\alpha\dot\alpha}$. 

The next question to ask is, what are the consequences of  central 
extension? In general, this question was investigated about 20 years 
ago 
\cite{OW}. It was shown that the nonvanishing central charge implies 
a lower bound on the mass of the soliton -- the domain wall in the 
case at hand. If the lower bound is saturated (see e.g. Eq. (\ref{wtsg})
in SUSY gluodynamics), the wall is  saturated; only 1/2 of 
supersymmetry is spontaneously broken on the wall,  the other half is 
preserved. Thus, if the wall is saturated, we know exactly its tension, 
although not much is known about its internal structure.

Assume that the walls in SUSY gluodynamics are saturated (I will 
argue that this is the case later). Then the wall tension is given by Eq. 
(\ref{wtsg}).  An intriguing question to study is
the dependence on the number of colors.
Apart from the explicit $N_c$ factor on the right-hand side, we 
should keep in mind an implicit $N_c$ dependence hidden in the 
gluino condensate. With my normalization of the fields, 
$\langle\mbox{Tr} \lambda^2\rangle
\sim N_c$; however, the variation of the gluino condensate
in the neighboring vacua is much smaller. Indeed, from 
Eq. (\ref{gc}) one infers that for the neighboring vacua
$\Delta\mbox{Tr} \lambda^2 \sim N_c^{-1} \mbox{Tr} 
\lambda^2\sim N_c^0$. This means that the tension of such walls 
scales as $\varepsilon\sim N_c^1$, a rather unexpected dependence. 
The scaling law typical of glueball  solitons is $N_c^2$ rather than 
$N_c$. 
Yet, the assumption of  BPS saturation, in conjunction with 
the expression for the central charge we derived, unambiguously 
leads us to the linear $N_c$ dependence. The very same linear 
dependence was  observed by Witten  who analyzed a theory  
related to SUSY  
gluodynamics from the point of
view of $D$-brane physics \cite{EW}. In this picture the domain 
walls
also appear naturally.  The linear $N_c$ dependenceÊ is natural from 
the $D$-brane 
perspective. The question is, can one understand the origin of the 
linear scaling law 
 in field theory {\em per se}, without any reference to  $D$-branes? 

The answer to this question is ``yes" \cite{KKS}. 
At large $N_c$, the mesons and the glueballs 
should have masses of order 1, trilinear couplings of
order $1/N_c$, and so on. This is conveniently 
encoded in an effective Lagrangian
of the form
\begin{equation}
{\cal L}=N^2_cF[M_i, \partial M_i]\, ,
\label{glueb}
\end{equation}
where $\{M_i\} $ is a set of fields (generically infinite)
representing  mesons and glueballs.
The value of the functional $F$ 
itself and all its derivatives at the minima should be
independent of $N_c$. This would ensure the proper $N_c$
dependence of the masses and coupling constants. 
Now, suppose we have a solution of classical equations of motion 
$M_{\rm wall}$
which describes
a wall
configuration interpolating between two distinct minima.
Since $N_c^2$ is an overall factor in Eq. (\ref{glueb}), 
at first  sight one may expect that $F[M_{\rm wall}]=O(1)$, 
and, therefore,   the wall tension
$\varepsilon=O(N_c^2)$. Such a situation is standard
in the soliton physics. 

How can one avoid this conclusion? Consider a function $F$ glued of 
$N_c$ sectors, so that it is non-analytic. 
At the minima the derivative of $F$ are $N_c$ independent. 
If at distances of order $1/N_c$ in the space of fields $M_i$ some 
of the derivatives become large, of order $N_c$ --
which is typical of glued $N_c$-sector functionals -- the naive $N_c$ 
counting is distorted, and the overall $N_c^2$ factor in Eq. 
(\ref{glueb}) is converted into $N_c$ in the soliton mass.

Thus, the $N_c$-sector structure inherent to SUSY gluodynamics is an 
implicit source of $N_c$ dependence. Of course, these sectors appear 
only in the ``macroscopic" glueball-based description of the theory
(\ref{glueb}). 
In the ``microscopic"  description everything is smooth. 
The fact that  the volume energy $E $ is
$O(N_c)$  inside the BPS wall 
connecting two neighboring vacua, say with Tr$\langle\lambda 
\lambda\rangle
=\Lambda^3 \exp (2\pi i k /N)$ where $k=0$ and 1, is seen in  the 
microscopic theory too.  Indeed, for the BPS wall
$$
G_{\mu\nu}^a G_{\mu\nu}^a \sim \partial_z \lambda^2
\sim \Delta \lambda^2 /L \, .
$$
 Since the volume 
energy density $E\sim N_c \, G_{\mu\nu}^a G_{\mu\nu}^a$,
and $\Delta \lambda^2$ in the neighboring chirally asymmetric 
vacua is $O(1)$, we conclude that $E$ scales as $N_c$.

Let me note in passing that the $D$-brane analysis \cite{EW}
suggests that the
confining QCD string emanating from the probe color  charges 
(quarks) on one side of the wall can
terminate on the wall,  without penetrating on 
the other side. In \cite{KKS} it is explained how this feature emerges
in the field-theoretic picture of the domain walls. 

Now, we pass to the most crucial point -- whether or not the
domain walls in SUSY gluodynamics are BPS saturated. This is a 
dynamical question.
{\em A priori} it is impossible to assert that they are  BPS saturated.
A part of the walls -- or all of them -- could be nonsaturated. Then,
Eq. (\ref{wtsg}) would play the role of a lower bound rather than the 
exact prediction. 

We have very few tools which might help us to answer this question 
in SUSY gluodynamics. One can try to apply the 
Veneziano-Yankielowicz effective Lagrangian \cite{veneziano}, 
amended in
\cite{KS}
to properly
incorporate 
the non-anomalous $Z_{N_c}$ symmetry of SUSY gluodynamics.
Of course, since this Lagrangian is not truly Wilsonean, one cannot
trust it in the quantitative aspect. However, since  saturation {\em 
versus} non-saturation is a qualitative  issue -- the answer 
Êdepends only on a global ``geography" in the space of fields --
it is reasonable to expect that the Veneziano-Yankielowicz 
Lagrangian can be applied for this limited purpose. 
Is this expectation  true or false? Both...

The  Lagrangian realizing all anomalous Ward
identities  is written in terms of 
the chiral superfield 
$
S = {3}/({32\pi^2N_c})\,\mbox{Tr}\,W^2$,
\beq
{\cal L} = \frac{1}{4}\int d^4\theta 
N^2_c\left( \bar S S \right)^{1/3} 
+ \frac{1}{3} \int d^2\theta N_c S\left( \ln {S^{N_c}} +2\pi i 
n\right) +\mbox{H.c.}
\label{VYL}
\eeq

From  this Lagrangian you  see that the potential
energy is built from $N_c$ distinct sectors,
\beq
U(\phi ) = 4N^2_c(\phi^*\phi)^{2/3}
\ln (\phi e^{-i{2\pi n/ N_c}}) \ln (\phi^*e^{i{2\pi n/ N_c}})
\label{Uphi}
\eeq
$$
\mbox{at}  \ \ \ \ {(2n-1)\pi\over N_c}< {\rm  arg}\phi
<{(2n+1)\pi\over N_c}\, .
$$
The $Z_{N_c}$ symmetry is explicit in this expression.
Including the kinetic term one gets the following   effective 
Lagrangian:
\beq
{\cal L}=N^2_c\{\partial_\mu 
\phi^{1/3}\partial_\mu\phi^{*1/3}+U(\phi)\}\, .
\label{effl}
\eeq
It is quite obvious that we deal here with $N_c+1$
supersymmetric minima -- $N_c$ minima  at $\phi = e^{i{2\pi n/ 
N_c}}$, 
corresponding 
to a non-vanishing value of the gluino condensate (spontaneously 
broken discrete chiral symmetry), and a minimum at $\phi = 0$
(unbroken chiral symmetry). 
The scalar potential itself is 
continuous,
but its first derivative in the angular direction experiences a jump
at arg$\phi = (2n+1)\pi/N_c$. 

Equations (\ref{Uphi}) and (\ref{effl}) do imply that
the wall connecting any of the chirally asymmetric vacua with the 
chirally symmetric one at the origin is  saturated. Building this wall 
is  more or less a straightforward exercise \cite{KSS}.
At the same time, the walls connecting 
the chirally asymmetric vacua with each other are not seen, at least 
at $N_c=3$ \cite{smilga}. In fact, they are not expected to be seen, 
since whenever a cusp is  crossed, the description based on 
the 
Veneziano-Yankielowicz 
Lagrangian fails \cite{KKS}. The restructuring of  degrees of 
freedom
responsible for the cusp is not properly reflected. 
Thus,  in this case, the existence of the BPS walls remains to 
be confirmed. Witten's $D$ brane construction \cite{EW}
can be viewed  as an indirect argument and  a guideline. 

Since the chirally symmetric vacuum plays
such an important role in this range of questions,
it is worth discussing  in more detail. 
Apart from the 
Veneziano-Yankielowicz 
Lagrangian there is one more argument (I call it the {\em missing 
state argument}) pointing in this direction. It is based on the 
comparison of two alternative calculations of 
$\langle\mbox{Tr}\lambda^2\rangle$.
One is the direct one-instanton calculation in the strong coupling 
regime. Another calculation uses the trick I  mentioned at the 
beginning of my talk -- the theory is expanded by adding matter 
fields which set up the Higgs regime. This is a weak coupling 
calculation, believed to be absolutely clean.
The two results do not match \cite{NSVZ}!  One can argue that
the instanton calculation in the strong coupling regime
averages the gluino condensate over all vacuum states \cite{amati}.
If so, the existence of the chirally symmetric vacuum
can be inferred through a leakage in this vacuum which would 
explain the mismatch.

The chirally symmetric vacuum at 
$\langle\mbox{Tr}\lambda^2\rangle =0$ is not seen in Witten's 
$D$-brane construction. Perhaps, this is not surprising at all. 
Indeed, there is a good deal of extrapolation in this construction, 
against which the chirally symmetric vacua are stable
(they have no choice since they have to saturate Witten's index)
while the $\langle\mbox{Tr}\lambda^2\rangle =0$ vacuum need not 
be stable. In fact, as I mentioned, it disappears in finite volume.
Evidently, it cannot survive the space-time distortions associated 
with the $D$-brane engineering. 

Concluding my talk, I would like to make two brief remarks
regarding the topics I  had no time to cover. Both remarks do not 
necessarily refer to gauge theories; they are of a more general 
nature.

First,  supersymmetric domain walls give a very interesting 
example of a
degeneracy which has no analogs in the nonsupersymmetric case.
Consider a theory with, say, three discrete vacua, such that the 
values of the superpotential in all vacua have one and the same 
phase.
(In the gauge theories one should consider a generalized 
superpotential, see Eq. (\ref{mlpcedop})). Such a situation is quite 
common in the Wess-Zumino models with two or more fields.
Then $C_{13} = C_{12} + C_{23}$ where $C$'s denote the 
corresponding central charges.
If BPS walls exist in all possible transitions, $12$, $13$ and $23$,
the wall tensions will obey a similar relation, see Eq. (\ref{degen}). 
This implies, in turn, that the $12$ and $23$ walls
at rest do not interact with each other no matter how large/small the 
distance is  between them \cite{MS,SV}.
In a two-field Wess-Zumino model, it is possible to find 
an analytic solution describing a superposition of the $12$ and $23$ 
walls,
for the first time ever \cite{SV}. 

Second, one can use  supersymmetric domain walls and similar 
topological defects in the cosmological aspect,
as a possible mechanism for breaking SUSY, in  part  or completely, 
at scales much lower than $M_{\rm Planck}$.
It is difficult to overestimate the value of possible alternative
scenarios of SUSY breaking occurring at a human scale.
The  topic  was open for discussion in Ref. 1 and was further 
elaborated in \cite{DD} where a relatively viable and even attractive 
scenario was put forward. 

\section*{Acknowledgments}

This work was supported in part by DOE under the grant number
DE-FG02-94ER40823.

\end{document}